\documentclass{article}
\usepackage{spconf,amsmath,graphicx}
\usepackage{booktabs}
\usepackage{caption}

\title{Generalization ability of MOS prediction networks}
\name{Erica Cooper$^1$, Wen-Chin Huang$^2$, Tomoki Toda$^2$, Junichi Yamagishi$^1$}
\address{ $^1$National Institute of Informatics, Japan \\
  $^2$Nagoya University, Japan}
  
\begin{document}
\ninept   
\maketitle
\begin{abstract}
Automatic methods to predict listener opinions of synthesized speech remain elusive since listeners, systems being evaluated, characteristics of the speech, and even the instructions given and the rating scale all vary from test to test.  While automatic predictors for metrics such as mean opinion score (MOS) can achieve high prediction accuracy on samples from the same test, they typically fail to generalize well to new listening test contexts.  In this paper, using a variety of networks for MOS prediction including MOSNet and self-supervised speech models such as wav2vec2, we investigate their performance on data from different listening tests in both zero-shot and fine-tuned settings.  We find that wav2vec2 models fine-tuned for MOS prediction have good generalization capability to out-of-domain data even for the most challenging case of utterance-level predictions in the zero-shot setting, and that fine-tuning to in-domain data can improve predictions.  We also observe that unseen systems are especially challenging for MOS prediction models. 
\end{abstract}
\begin{keywords}
Speech synthesis, mean opinion score, speech naturalness assessment, MOS prediction
\end{keywords}
\section{Introduction}
\label{sec:intro}

Listening tests with human subjects are the gold standard for evaluating synthesized speech, but these tests can take a long time and become cost-prohibitive as the number of systems to evaluate increases.  Automatic mean opinion score (MOS) prediction would enable faster experimental iteration as well as larger-scale experiments, but this technology has a long way to go.  Every set of systems or samples in a listening test comprises a unique context, with different listeners, ranges of systems being evaluated, and even instructions.  Thus, predicting MOS using a model pretrained on one listening test typically does not generalize well to others.  Can we design MOS prediction models that have better generalization abilities?   Can generalizable MOS prediction models be utilized on a new listening test context in a zero-shot manner, or is fine-tuning necessary?

In an initial step towards answering these questions, we use a dataset of diverse synthesized speech samples and their MOS ratings that we have previously collected in a large-scale listening test for this purpose \cite{ssw11mos}.  In this work, we design training, development, and test set splits for this data  such that the development and test sets contain unseen speakers, systems, listeners, and texts, in order to stress-test MOS prediction networks with challenging cases, and to investigate which of these factors affect MOS prediction performance.  We also gather additional ``out-of-domain'' datasets to study the generalization ability of MOS predictors.  We explore severall different model types and configurations, including original MOSNet \cite{mosnet} and finetuning large-scale self-supervised speech models \cite{wav2vec2,hubert}.

\vspace{-1mm}
\section{Related Work}
\label{sec:relatedwork}
\vspace{-1mm}

Automatic MOS prediction using neural networks has become a research topic of interest.  One such investigation is MOSNet \cite{mosnet}, which uses a CNN-BLSTM architecture to predict naturalness ratings of voice conversion samples from their magnitude spectrograms.  An extension of this work in \cite{williams2020comparison} investigated different input feature representations such as speech embeddings.  Considering the large variations in listener preferences, one popular approach is to explicitly model the listener dependencies of MOS scores, as in MBNet \cite{leng2021mbnet}, which uses listener labels during training as input to a listener-bias branch of the model, and \cite{tseng2021utilizing}, which learns a listener bias during the fine-tuning of large-scale self-supervised speech models for the MOS prediction task.  One common theme in these works is that utterance-level ratings are more difficult to predict than system-level ones.  Another theme in these papers is that these models tend not to generalize well to data from other listening tests.  In this work, we investigate different types of networks for MOS prediction, and aim to better understand their generalization capability and the conditions in which they can be successful at predicting MOS for unseen data and different listening test contexts.


\vspace{-1mm}
\section{Datasets}
\label{sec:datasets}
\vspace{-1mm}

We make use of one main training dataset based on a listening test that we previously conducted on combined samples from  many different systems from past years going back to 2008, as well as three additional ``out-of-domain'' datasets from past listening tests.  In constructing training, development, and test sets, we aimed to match the distributions of the averaged MOS of the samples in each set to the overall distribution, and furthermore, to match the distributions of {\em standard deviations} of ratings per utterance, since we found in our prior work that some systems were more ``controversial'' than others, with a wide distribution of scores.  We also required that both development and test sets should have unseen speakers, systems, listeners, and texts, wherever possible.  

To create one candidate training/development/test split, we chose without replacement some unseen speakers,  systems,  texts, and  listeners for each of the development and test sets.  Unseen categories in the development set are unseen with respect to the training set, and unseen categories in the test set are unseen with respect to {\em both} the training and development sets.  The target number of audio samples per set is then filled by randomly selecting from the remaining utterances.  We evaluated a candidate split by earth-mover's distance (EMD) between the distribution of the total data and each subset:  the evaluation metric was the sum of EMD for individual scores for the training, development, and test set, plus the EMD for {\em standard deviations} of each set, as compared to the full data.  We iterated this random sampling to create candidate splits 1000 times with different random seeds, and picked the one with the lowest sum of EMDs (a lower EMD value indicates that the distribution of each subset is close to the distribution of the overall data, and that therefore the split is well-balanced).  All audio files were downsampled to 16kHz to match the lowest sampling rate.

Descriptions of each dataset follow; a summary is in Table \ref{tab:data}.

\vspace{-2mm}
\addtolength{\tabcolsep}{-3pt}
\begin{table}[th]
\scriptsize

  \caption{Datasets: audio samples, ratings per sample, speakers, and systems,  and unseen categories per development and test set.}
  \label{tab:data}
  \centering
    \vspace{-2mm}
  \begin{tabular}{ l l l l l l l l l }
    \toprule
    Name & samp & ratings & spk & sys & unseen  & unseen & unseen  & unseen  \\
    & & per samp & &  & spk &   sys & listeners & texts \\
    \midrule
    BVCC & 7106 & 8 & 27 & 187 & 1 & 6 & 8 & 5 \\
    ASV2019 & 18079 & 1-26 & 67 & 14 & 4 & 2 & 10 & - \\
    BC2019 & 1352 & 10-17 & 1 & 26 & - & 2 & 70 & 2 \\
    COM2018 & 4760 & 1-9 & 1 & 10 & - & 1 & 5 & 5 \\
    \bottomrule
  \end{tabular}
\vspace{-2mm}
\end{table}
\addtolength{\tabcolsep}{3pt}

\vspace{-4mm}
\subsection{In-domain data}

\textbf{BVCC}  We conducted a large-scale listening test on samples from past speech synthesis challenges and open-source implementations, the results of which we published in \cite{ssw11mos}; we name this dataset BVCC since most samples are from the \textbf{B}lizzard Challenge for TTS and the \textbf{V}oice \textbf{C}onversion \textbf{C}hallenge.  We focused on English-language synthesis and the main Hub tasks for each year.  The Blizzard Challenges that we included were \cite{blizzard2008,blizzard2009,blizzard2010,blizzard2011,blizzard2013,blizzard2016}, as well as all Voice Conversion Challenge years  \cite{vcc2016description,vcc2016analysis,vcc2018,vcc2020,Rohan2020}.  We also included publicly-available samples from systems implemented in ESPnet \cite{watanabe2018espnet}, a popular open-source toolkit for end-to-end speech technologies \cite{hayashi2019espnettts}.  
We re-evaluated all of these samples in one listening test in order to create one unified listening test context for this large variety of samples -- otherwise, samples from different tests are not directly comparable, since they come from different contexts.  We created a training/development/test split of  70\%/15\%/15\%.\footnote{We plan to publicly release this dataset and its splits in the near future.}

\vspace{-2mm}
\subsection{Out-of-domain data}

For out-of-domain data, we made use of various archives of past listening tests and their original ratings; no new listening tests were conducted.  We consider these datasets to be out-of-domain because they come from different listening tests with different ranges of sample quality,  listeners, and instructions.  We looked at the ASVSpoof 2019 Logical Access (LA) samples and their listening test ratings \cite{WANG2020101114,Todisco2019}, the Blizzard Challenge 2019 listening test data \cite{wu2019blizzard}, and a listening test from 2018 comparing various combinations of acoustic models and vocoders \cite{wang2018comparison}.  We created fine-tuning/development/test splits of 33\%/33\%/33\% for each of these databases; we choose a smaller fine-tuning proportion because this data is intended to fine-tune models which have already seen the larger BVCC training data, and is meant to represent a condition where a small amount of data from a target listening test context is available.  This out-of-domain data will only be used for for fine-tuning models that have already been trained (or fine-tuned) on BVCC, and for testing.

\textbf{ASV2019}  English synthesized audio samples from a variety of state-of-the-art speech synthesis and voice conversion systems prepared for the ASVSpoof Challenge in 2019, in which participants submit systems to detect spoofed vs.~bona fide audio.  In the listening test, listeners judged whether a sample was produced by a machine or a human on a scale from 1-10, where 1 is definitely machine generated and 10 is definitely human; we linearly adjusted these scores to our standard scale of 1-5.  Most audio samples only have one rating, and natural audio is over-sampled and has up to 26 ratings because the aim of this listening test was to measure human performance on spoofing detection as compared to automatic detection, rather than to evaluate the quality of different synthesis methods.  The different target task of this listening test creates a challenging domain mismatch.  We did not include standard deviations in the EMD sum metric since most samples only had one rating.

\textbf{BC2019}  Chinese TTS samples submitted to the 2019 Blizzard Challenge, rated by native speakers of Chinese.  Since all of the BVCC samples are English, data in a different language is a challenging domain-mismatched condition which will allow us to study whether MOS predictors can generalize well across languages.

\textbf{COM2018}  This listening test was a comparison of 9 different combinations of four acoustic models and four  vocoders, plus natural speech, using data from the Japanese female speaker ``F009'' from the XIMERA database \cite{kawai2004ximera}, as another cross-language condition.  

\vspace{-2mm}
\subsection{Data distributions}

\begin{figure}[t!] 
\centering
\includegraphics[width=1.\linewidth]{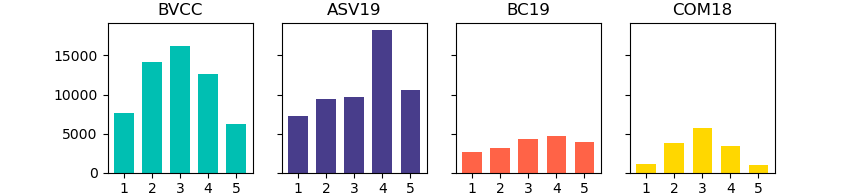}
\vspace{-6mm}
\caption{Distributions of scores for each dataset}
\label{fig:distr}
\vspace{-6mm}
\end{figure}

Each dataset has a different distribution of scores due to the differing nature and context of each listening test, as illustrated in Figure \ref{fig:distr}, which shows the number of ratings for each score.  Adjusted ASV2019 scores were rounded to the nearest integer for clarity.

\vspace{-2mm}
\section{Experiments and Results}
\label{sec:experiments}
\vspace{-1mm}

We conduct experiments using original MOSNet \cite{mosnet}, as well as various self-supervised-learning-based (SSL) speech models from the Fairseq\footnote{https://github.com/pytorch/fairseq} project, which have shown to be useful via fine-tuning for diverse speech tasks, such as phoneme recognition, speaker identification, spoken language understanding, and emotion recognition \cite{yang2021superb}.  A summary of the publicly-available Fairseq models (wav2vec2 \cite{wav2vec2} and HuBERT \cite{hubert}) that we investigated is in Table \ref{tab:fairseq}.

 \addtolength{\tabcolsep}{-2pt}
 \begin{table}[th]
 \vspace{-1mm}
\footnotesize
  \caption{Information about Fairseq pretrained base models}
  \label{tab:fairseq}
    \vspace{-2mm}
  \centering
  \begin{tabular}{ l  l  l  l}
    \toprule
    Name & Training data & \# params & Out dim. \\
    \midrule
    \multicolumn{4}{c}{wav2vec2}  \\
    \midrule 
    w2v\_small  & Librispeech \cite{librispeech} & 95m & 768 \\
    libri960\_big   & Librispeech & 317m & 1024 \\
    w2v\_vox\_new & Libri-Light \cite{librilight} & 317m & 1024 \\
    w2v\_large & Libri-Light,  & 317m & 1024 \\
     & CommonVoice \cite{commonvoice}, \\
     &  Switchboard \cite{switchboard}, Fisher \cite{fisher}  \\
    xlsr & MLS \cite{mls}, CommonVoice,  & 317m & 1024 \\
     & BABEL \cite{babel} \\
    \midrule
    \multicolumn{4}{c}{HuBERT} \\
    \midrule
    hubert\_base\_ls960 & Librispeech & 95m & 768 \\
    hubert\_large\_ll60k & Libri-Light & 316m & 1024 \\
    \bottomrule
  \end{tabular}
  \vspace{-4mm}
\end{table}
\addtolength{\tabcolsep}{2pt}

In addition to mean squared error (MSE), we also consider various correlation metrics since it is also important for the relative orderings of the scores to be predicted correctly.  We thus also report Linear Correlation Coefficient (LCC) as a basic correlation measure, Spearman Rank Correlation Coefficient (SRCC) which is  non-parametric and  measures correlation of ranking order, and Kendall Tau Rank Correlation (KTAU), another type of rank correlation which tends to be more robust to errors.
 

\begin{figure*}[ht]
\centering
\includegraphics[width=1.\linewidth]{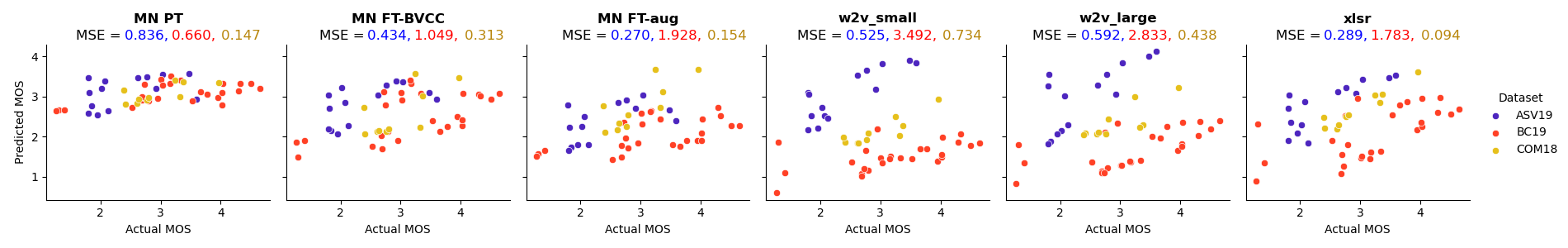}
\vspace{-8mm}
\caption{Scatter plot of system-level zero-shot prediction results for each system.
}
\vspace{-3mm}
\label{fig:scatter}
\end{figure*}

\begin{figure*}[ht]
\centering
\includegraphics[width=1.\linewidth]{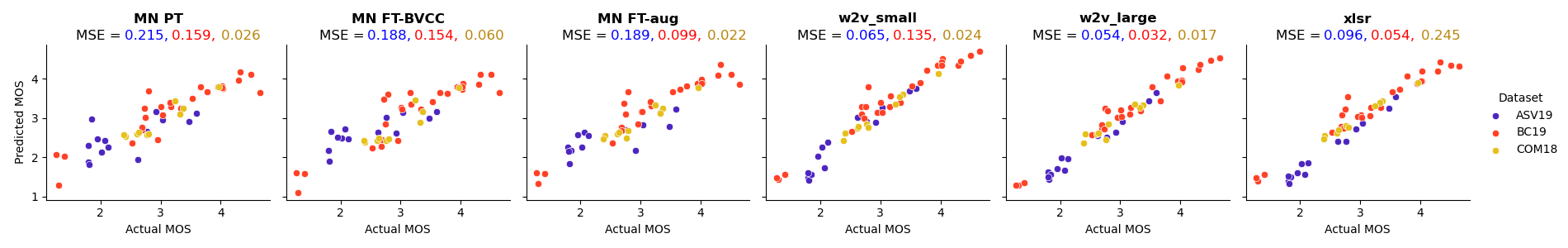}
\vspace{-8mm}
\caption{Scatter plot of system-level fine-tune prediction results for each system.
}
\vspace{-4mm}
\label{fig:scatter_ft}
\end{figure*}

\vspace{-3mm}
\subsection{MOSNet}

We first investigate the original MOSNet \cite{mosnet} CNN-BLSTM architecture trained from scratch on BVCC.  We also try fine-tuning the pretrained model provided by the authors.  We also explore two  data augmentation strategies: perturbing the audio speed by a randomly-chosen factor between 0.95 and 1.05, 
and trimming or adding silence by a small random value. 
We run speedup, slowdown, trimming, and adding silence on the entire dataset,  resulting in a total of 5 times the original data when all augmentations are used.  We also evaluated the pretrained MOSNet in a zero-shot manner without any fine-tuning.  Since the pretrained model was trained on VCC2018, samples from this challenge are not unseen, so we exclude these from our development and test sets for all experiments.  Test set results are in Table \ref{tab:mosnetresults}; best results for each evaluation metric are in bold.

\addtolength{\tabcolsep}{-2.5pt}
\begin{table}[th]
\vspace{-1mm}
\scriptsize
  \caption{MOSNet BVCC results}
  \label{tab:mosnetresults}
    \vspace{-2mm}
  \centering
  \begin{tabular}{ l | l l l l | l l l l}
    \toprule
    &  \multicolumn{4}{c}{\textbf{Utterance level}} & \multicolumn{4}{c}{\textbf{System level}} \\
    Model &  MSE & LCC & SRCC & KTAU & MSE & LCC & SRCC & KTAU \\
    \midrule
    Pretrained \cite{mosnet} & 0.831 & 0.374 &	0.393 &	0.275 &	0.541 &	0.354 &	0.352 &	0.243  \\
    From scratch & 0.777 &	0.304 &	0.261 &	0.178 &	0.504 &	0.239 &	0.181 &	0.117 \\
    Fine-tuned & 0.417 & 0.715 & 0.711 &	0.529 &	0.162 &	0.852 &	0.862 &	0.663 \\ 
    FT$+$sil.aug & 0.428 & 0.713 & 0.709 &	0.528 &	0.153 &	0.854 &	0.861 &	0.665 \\
    FT$+$speed aug & 0.421 &	0.716 &	0.707 &	0.526 &	0.176 &	0.857 &	0.867 &	0.672 \\
    FT$+$both aug & \textbf{0.305} & \textbf{0.796} & \textbf{0.791} & \textbf{0.604} & \textbf{0.096} & \textbf{0.905} & \textbf{0.912} &	\textbf{0.737} \\    
    \bottomrule
  \end{tabular}
  \vspace{-1mm}
\end{table}
 \addtolength{\tabcolsep}{2.5pt}

Surprisingly, we found that training from scratch on BVCC was worse than simply using the pretrained model.  This may be because although our BVCC listening test was large in scale and covered a large variety of systems, the number of audio files in the training data is much smaller (4974, as compared to 13580 in the VCC2018 training set); even though our dataset has more ratings per sample, the {\em averaged} ratings are used for training and evaluation.  Our dataset may simply not contain enough examples to train MOSNet from scratch.  Fortunately, we find that fine-tuning the pretrained model on BVCC gives a large jump in performance, and furthermore, fine-tuning on all types of augmented data gives further improvements.

\vspace{-4mm}
\subsection{Fairseq}
\vspace{-1mm}

The strong performance of fine-tuned speech SSL models on diverse tasks motivates us to try this approach for MOS prediction.
We fine-tune various wav2vec2 and HuBERT pretrained SSL models by mean-pooling the model's output embeddings, adding a linear output layer, and training with L1 loss.\footnote{Implementation: https://github.com/nii-yamagishilab/mos-finetune-ssl}  This is a similar approach to \cite{tseng2021utilizing}, who also fine-tuned SSL models for the MOS prediction task, but our aims are different: while the authors modeled listener differences, our purpose is to investigate the generalization capabilities of different base models using very simple fine-tuning.    
We found in preliminary experiments that including augmented data during fine-tuning did not improve the MOS prediction results of SSL models.  
Results of fine-tuning each base model on the training set of BVCC, and evaluating on the BVCC test set, can be seen in Table \ref{tab:fairseqresults}.

 \addtolength{\tabcolsep}{-3pt}
\begin{table}[th]
\vspace{-2mm}
\scriptsize
  \caption{Fine-tuned Fairseq BVCC results}
  \label{tab:fairseqresults}
  \vspace{-2mm}
  \centering
  \begin{tabular}{ l | l l l l l l l l}
    \toprule
    &  \multicolumn{8}{c}{\textbf{Test set}} \\
    &  \multicolumn{4}{c}{\textbf{Utterance level}} & \multicolumn{4}{c}{\textbf{System level}} \\
    Base model & MSE & LCC & SRCC & KTAU & MSE & LCC & SRCC & KTAU \\
    \midrule
    w2v\_small & 0.227 & \textbf{0.868} & \textbf{0.866} & \textbf{0.690} & 0.121 & 0.938 & 0.942 & 0.790 \\
    libri960\_big & 0.342 & 0.823 & 0.820 & 0.635 & 0.136 & 0.901 & 0.901 & 0.730 \\
    w2v\_vox\_new & 0.342 & 0.767 & 0.753 & 0.570 & 0.112 & 0.903 & 0.900 & 0.721 \\
    w2v\_large & \textbf{0.220} & \textbf{0.868} & 0.865 & \textbf{0.690} & \textbf{0.059} & \textbf{0.948} & \textbf{0.944} & \textbf{0.803} \\
    xlsr\_53\_56k & 0.281 & 0.821 & 0.816 & 0.633 & 0.107 & 0.902 & 0.894 & 0.730 \\
    hubert\_base\_ls960 & 0.318 & 0.842 & 0.837 & 0.655 & 0.213 & 0.919 & 0.915 & 0.745 \\
    hubert\_large\_ll60k & 0.444 & 0.696 & 0.687 & 0.507 & 0.184 & 0.812 & 0.805 & 0.620 \\
    \bottomrule
  \end{tabular}
  \vspace{-2mm}
\end{table}
 \addtolength{\tabcolsep}{3pt}

We observe that the best results are consistently from the (relatively) small wav2vec2 model and the large wav2vec2 model trained on a variety of different speech corpora.  The wav2vec2 model trained on multilingual data also had the third-best performance on the development set.  

\vspace{-3mm}
\subsection{Out-of-domain data experiments}
\vspace{-1mm}

We picked the best and most interesting models from the previous two experiments and tried both zero-shot MOS prediction on our three different out-of-domain datasets, and also fine-tuning on each dataset, in order to study generalization ability.  We consider the MOSNet pretrained on VCC2018 (MN PT), the pretrained MOSNet fine-tuned to our BVCC data (MN FT-BVCC), the fine-tuned MOSNet including all augmented data (MN FT$+$aug), and the best three wav2vec2 models, which also happen to cover an interesting variety of these models: a (relatively) small English-trained model, a large English model, and a large multilingual model.  We hypothesize that the multilingual model may generalize better to different languages such as Chinese and Japanese.

For the zero-shot condition, we simply use our existing models to make predictions on each of the out-of-domain test sets.  For the fine-tuning condition, we fine-tune each model using the fine-tuning portion of one dataset, and evaluate on that same dataset's test portion.  The fine-tuning condition represents a scenario where a small amount of listening test data is available or can be collected for a particular listening test context.  Note that some models will have been fine-tuned twice, first on the BVCC data and then on one out-of-domain set.  Zero-shot and fine-tuning results on each test set at the utterance level can be found in Table \ref{tab:ood}; system-level results are shown in the scatter plots in Figure \ref{fig:scatter} and Figure \ref{fig:scatter_ft}.

 \addtolength{\tabcolsep}{-2.5pt}
\begin{table}[th]
\vspace{-1mm}
\scriptsize
  \caption{Out-of-domain utterance-level results}
  \label{tab:ood}
  \centering
    \vspace{-2mm}
  \begin{tabular}{ l | l l l l | l l l l }
    \toprule
    & \multicolumn{4}{c}{\textbf{Zero-shot}} & \multicolumn{4}{c}{\textbf{Fine-tune}} \\
    Model & MSE & LCC & SRCC & KTAU & MSE & LCC & SRCC & KTAU  \\
    \midrule
    \multicolumn{9}{c}{\textbf{ASV2019}} \\
    \midrule
    MN PT & 1.912 & 0.142 & 0.159 & 0.112 &  1.217 & 0.379 & 0.386 & 0.273  \\
    MN FT-BVCC & 1.641 & 0.218 & 0.219 & 0.154 & 1.249 & 0.386 & 0.401 & 0.286  \\
    MN FT$+$aug & 1.617 & 0.199 & 0.218 & 0.153 & 1.240 & 0.368 & 0.377 & 0.268 \\
    w2v\_small & 1.498 & \textbf{0.470} & \textbf{0.491} & \textbf{0.352} & 1.073 & 0.541 & \textbf{0.558} & \textbf{0.405}  \\
    w2v\_large & 1.589 & 0.453 & 0.478 & 0.344 & \textbf{1.065} & \textbf{0.548} & 0.557 & 0.404  \\
    xlsr & \textbf{1.371} & 0.409 & 0.423 & 0.301 & 1.192 & 0.518 & 0.525 & 0.377 \\
    \midrule
    \multicolumn{9}{c}{\textbf{BC2019}} \\
    \midrule
    MN PT & \textbf{0.823} & 0.432 & 0.402 & 0.276 & 0.443 & 0.738 & 0.690 & 0.514 \\
    MN FT-BVCC & 1.328 & 0.444 & 0.470 & 0.321 & 0.444 & 0.743 & 0.692 & 0.517 \\
    MN FT$+$aug & 2.202 & 0.407 & 0.488 & 0.334 & 0.406 & 0.770 & 0.705 & 0.526 \\
    w2v\_small & 3.672 & 0.553 & 0.559 & 0.409 & 0.356 & 0.878 & 0.840  & 0.651 \\
    w2v\_large & 3.023 & 0.575 & \textbf{0.618} & \textbf{0.440} & \textbf{0.235} & \textbf{0.879} & \textbf{0.841} & \textbf{0.653} \\
    xlsr & 1.924 & \textbf{0.576} & 0.596 & 0.414 & 0.274 & 0.858 & 0.812 & 0.621 \\
    \midrule
    \multicolumn{9}{c}{\textbf{COM2018}} \\
    \midrule
    MN PT & \textbf{0.510} & 0.398 & 0.383 & 0.269 & 0.404 & 0.574 & 0.533 & 0.386 \\
    MN FT-BVCC & 0.768 & 0.420 & 0.391 & 0.276 & 0.458 & 0.558 & 0.535 & 0.387 \\
    MN FT$+$aug & 0.797 & 0.375 & 0.357 & 0.251 & 0.433 & 0.550 & 0.522 & 0.376 \\
    w2v\_small & 1.200 & 0.476 & 0.423 & 0.297 & \textbf{0.352} & \textbf{0.674} & \textbf{0.667} & \textbf{0.497} \\
    w2v\_large & 0.951 & 0.425 & 0.380 & 0.268 & 0.436 & 0.559 & 0.535 & 0.387 \\
    xlsr & 0.558 & \textbf{0.501} & \textbf{0.480} & \textbf{0.341} & 1.383 & 0.369 & 0.379 & 0.268  \\
    \bottomrule
  \end{tabular}
  \vspace{-3mm}
\end{table}
 \addtolength{\tabcolsep}{2.5pt}

As expected, the zero-shot condition is more challenging than fine-tuning.  We also observe the effect of number of ratings per utterance -- for ASV2019, for which many utterances have only one rating, we observe overall worse performance, even in the fine-tuning condition, reflecting the unpredictability of listener differences.  We also observe that the best-correlated model for the Japanese data for the zero-shot condition was the multilingual `xlsr' model, however this was not the case for the Chinese data.  For all datasets, wav2vec2 models demonstrated good generalizability, even in the challenging zero-shot scenario.  Although interestingly MOSNet models sometimes had the lowest MSE, wav2vec2 models consistently outperformed them in correlations.  In fact, despite the challenging nature of zero-shot prediction of utterance-level scores as compared to the fine-tuning setting or system-level predictions, wav2vec2 models are able to reach moderate correlations for this task.

Scatter plots of the system-level zero-shot results can be found in Figure \ref{fig:scatter}.  We observe that original pretrained MOSNet tends to restrict predictions to a narrow range, fine-tuning with additional BVCC data improves on that slightly, and Fairseq models improve further; these tend to under-predict scores for BC2018 and over-predict ASV2019, but less so in the case of multilingual xlsr.

Fine-tuning on a small amount of in-domain data reduces error rates and improves correlations, both at the utterance level (Table \ref{tab:ood}) and at the system level, as shown in the scatter plots in Figure \ref{fig:scatter_ft}.  Fine-tuning appears to mitigate MOSNet's tendency to predict only within a certain range, but the wav2vec2 models appear to benefit even more from fine-tuning.  The multilingual xlsr model no longer has an advantage when fine-tuned, with the small or large English-trained wav2vec models having the best performance in all cases.  

Since we held out unseen speakers, systems, listeners, and texts, we further analyzed the fine-tuned systems to learn which unseen categories are most challenging.  For each of the utterance-level predicted results, we measured its squared error with respect to the actual MOS.  Then, we checked whether the utterance is from a seen or unseen category, and gathered the squared errors accordingly, i.e. one list of squared errors for seen speakers of the ASV2019 dataset, and one for unseen speakers.  Then, we conducted a two-sided t-test to determine whether the distributions of errors were significantly different at a level of $p\le 0.05$.  When the unseen category's mean squared  error is higher and the difference is significant, this indicates that the unseen category is more challenging to predict.  Since a given utterance may be rated by a mix of both seen and unseen listeners, we consider unseen listeners only for ASV2019, for which most utterances only had one rater.  Results are in Table \ref{tab:unseen}.




 \addtolength{\tabcolsep}{-3.5pt}
\begin{table}[th]
\vspace{-1mm}
\scriptsize
  \caption{Analysis of unseen categories.  Mean and standard deviations of squared errors for the unseen categories are shown.  Unseen categories whose mean squared error is significantly higher than their seen counterparts are shown in bold.}
  \label{tab:unseen}
  \centering
    \vspace{-2mm}
  \begin{tabular}{ l | l l l l l l }
    \toprule

    Data & MN PT & MN FT & MN FT-aug & w2v\_sm & w2v\_lg & xlsr  \\
    \midrule
    \multicolumn{7}{c}{\textbf{Unseen speakers}} \\
    \midrule
    ASV19 & \textbf{1.33$\pm$1.65} & 1.28$\pm$1.52 & 1.23$\pm$1.48 & 1.02$\pm$1.72 & 1.04$\pm$1.77 & 1.18$\pm$2.04 \\
    \midrule
    \multicolumn{7}{c}{\textbf{Unseen systems}} \\
    \midrule
    ASV19 & \textbf{1.36$\pm$1.45} & \textbf{1.43$\pm$1.51} & \textbf{1.43$\pm$1.54} & \textbf{1.23$\pm$1.58} & \textbf{1.26$\pm$1.82} & \textbf{1.43$\pm$2.15} \\
    BC19 & \textbf{0.77$\pm$1.11} & \textbf{0.67$\pm$1.04} & \textbf{0.76$\pm$1.10} & \textbf{0.87$\pm$0.98} & \textbf{0.41$\pm$0.61} & \textbf{0.56$\pm$0.78} \\
    COM18 & 0.42$\pm$0.61 & 0.50$\pm$0.71 & 0.47$\pm$0.68 & 0.33$\pm$0.48 & \textbf{0.52$\pm$0.74} & 0.35$\pm$0.51 \\
    \midrule
    \multicolumn{7}{c}{\textbf{Unseen listeners}} \\
    \midrule
    ASV19 & 0.76$\pm$1.13 & 0.70$\pm$1.19 & \textbf{0.71$\pm$1.25} & \textbf{0.58$\pm$1.46} & 0.55$\pm$1.55 & 0.57$\pm$1.62 \\
    \midrule
    \multicolumn{7}{c}{\textbf{Unseen texts}} \\
    \midrule
    BC19 & 0.30$\pm$0.31 & 0.26$\pm$0.36 & 0.35$\pm$0.52 & 0.26$\pm$0.43 & 0.13$\pm$0.16 & 0.23$\pm$0.40 \\
    COM19 & 0.43$\pm$0.69 & 0.51$\pm$0.82 & 0.48$\pm$0.76 & \textbf{0.47$\pm$0.71} & 0.49$\pm$0.75 & \textbf{0.51$\pm$0.78} \\
    \bottomrule
  \end{tabular}
  \vspace{-2mm}
\end{table}
 \addtolength{\tabcolsep}{3.5pt}
 
 For ASV2019 and BC2019, unseen systems were always significantly different; for COM2018 they were usually not -- this is likely because a ``system'' for COM2018 is a combination of acoustic model and vocoder, both of which have been seen in other combinations during training.  For unseen texts, most differences are not significant, except for the COM2018 dataset with two of the Fairseq models.  These models were originally developed for ASR, so they may be learning something about the text content of the utterances.

\vspace{-2mm}
\section{Conclusions and Future Work}
\label{sec:conclusions}
\vspace{-1mm}

We have shown that fine-tuning SSL models can enable MOS prediction for a new listening test context using a smaller amount of human-labeled MOS data, which is costly to obtain, than training a model for this purpose from scratch.  We found that MOSNets need a large amount of data for training from scratch, whereas fine-tuning is an effective way to make use of smaller datasets.  Large SSL models can be successfully used for the MOS prediction task, and they demonstrate good performance.  This is especially the case when target listening test data is available for fine-tuning, but these models can surprisingly do moderately well in even the very challenging case of zero-shot utterance-level prediction.  SSL models trained on multilingual data or on a mix of different datasets especially show good generalization ability.  

We have also identified the difficult cases for MOS prediction, which indicate the most interesting directions for future work.  Although prediction on unseen systems is a likely real-world use case for MOS predictors, this category remains the most challenging to predict.  


\vspace{2mm}

\noindent
\footnotesize{
\textbf{Acknowledgments} This study is supported by JST CREST grants JPMJCR18A6, JPMJCR20D3, and JPMJCR19A3, and by MEXT KAKENHI grants 21K11951 and 21K19808.  Thanks to the organizers of the Blizzard Challenge and Voice Conversion Challenge, and to Zhenhua Ling, Zhihang Xie, and Zhizheng Wu for answering our questions about past challenges.
}

\vfill\pagebreak

\footnotesize{

\bibliographystyle{IEEEbib}
\bibliography{strings,refs}

\begin{thebibliography}{10}

\bibitem{ssw11mos}
Erica Cooper and Junichi Yamagishi,
\newblock ``How do voices from past speech synthesis challenges compare
  today?,''
\newblock {\em Proceedings of the 11th ISCA Speech Synthesis Workshop}, 2021.

\bibitem{mosnet}
Chen-Chou Lo, Szu-Wei Fu, Wen-Chin Huang, Xin Wang, Junichi Yamagishi, Yu~Tsao,
  and Hsin-Min Wang,
\newblock ``{MOSNet:} deep learning-based objective assessment for voice
  conversion,''
\newblock in {\em Proc. Interspeech 2019}, 2019, pp. 1541--1545.

\bibitem{wav2vec2}
Alexei Baevski, Yuhao Zhou, Abdelrahman Mohamed, and Michael Auli,
\newblock ``wav2vec 2.0: A framework for self-supervised learning of speech
  representations,''
\newblock {\em Advances in Neural Information Processing Systems}, vol. 33,
  2020.

\bibitem{hubert}
Wei-Ning Hsu, Benjamin Bolte, Yao-Hung~Hubert Tsai, Kushal Lakhotia, Ruslan
  Salakhutdinov, and Abdelrahman Mohamed,
\newblock ``{HuBERT:} self-supervised speech representation learning by masked
  prediction of hidden units,''
\newblock {\em arXiv e-prints}, pp. arXiv--2106, 2021.

\bibitem{williams2020comparison}
Jennifer Williams, Joanna Rownicka, Pilar Oplustil, and Simon King,
\newblock ``Comparison of speech representations for automatic quality
  estimation in multi-speaker text-to-speech synthesis,''
\newblock {\em Speaker Odyssey}, 2020.

\bibitem{leng2021mbnet}
Yichong Leng, Xu~Tan, Sheng Zhao, Frank Soong, Xiang-Yang Li, and Tao Qin,
\newblock ``{MBNET: MOS} prediction for synthesized speech with mean-bias
  network,''
\newblock in {\em ICASSP 2021-2021 IEEE International Conference on Acoustics,
  Speech and Signal Processing (ICASSP)}. IEEE, 2021, pp. 391--395.

\bibitem{tseng2021utilizing}
Wei-Cheng Tseng, Chien-yu Huang, Wei-Tsung Kao, Yist~Y Lin, and Hung-yi Lee,
\newblock ``Utilizing self-supervised representations for {MOS} prediction,''
\newblock {\em Interspeech}, 2021.

\bibitem{blizzard2008}
Vasilis Karaiskos, Simon King, Robert~AJ Clark, and Catherine Mayo,
\newblock ``The {Blizzard Challenge} 2008,''
\newblock in {\em Proc. Blizzard Challenge Workshop}. Citeseer, 2008.

\bibitem{blizzard2009}
Alan~W Black, Simon King, and Keiichi Tokuda,
\newblock ``The {Blizzard Challenge} 2009,''
\newblock in {\em Proc. Blizzard Challenge}, 2009, pp. 1--24.

\bibitem{blizzard2010}
Simon King and Vasilis Karaiskos,
\newblock ``The {Blizzard Challenge} 2010,''
\newblock 2010.

\bibitem{blizzard2011}
Simon King and Vasilis Karaiskos,
\newblock ``The {Blizzard Challenge} 2011,''
\newblock 2011.

\bibitem{blizzard2013}
Simon King and Vasilis Karaiskos,
\newblock ``The {Blizzard Challenge} 2013,''
\newblock 2013.

\bibitem{blizzard2016}
Simon King and Vasilis Karaiskos,
\newblock ``The {Blizzard Challenge} 2016,''
\newblock 2016.

\bibitem{vcc2016description}
Tomoki Toda, Ling-Hui Chen, Daisuke Saito, Fernando Villavicencio, Mirjam
  Wester, Zhizheng Wu, and Junichi Yamagishi,
\newblock ``The {Voice Conversion Challenge} 2016,''
\newblock in {\em Interspeech}, 2016.

\bibitem{vcc2016analysis}
Mirjam Wester, Zhizheng Wu, and Junichi Yamagishi,
\newblock ``Analysis of the {Voice Conversion Challenge} 2016 evaluation
  results.,''
\newblock in {\em Interspeech}, 2016, pp. 1637--1641.

\bibitem{vcc2018}
Jaime Lorenzo-Trueba, Junichi Yamagishi, Tomoki Toda, Daisuke Saito, Fernando
  Villavicencio, Tomi Kinnunen, and Zhenhua Ling,
\newblock ``The {Voice Conversion Challenge} 2018: Promoting development of
  parallel and nonparallel methods,''
\newblock .

\bibitem{vcc2020}
Zhao Yi, Wen-Chin Huang, Xiaohai Tian, Junichi Yamagishi, Rohan~Kumar Das, Tomi
  Kinnunen, Zhenhua Ling, and Tomoki Toda,
\newblock ``{Voice Conversion Challenge} 2020 — intra-lingual semi-parallel
  and cross-lingual voice conversion —,''
\newblock in {\em Proc. Joint Workshop for the Blizzard Challenge and Voice
  Conversion Challenge 2020}, 2020, pp. 80--98.

\bibitem{Rohan2020}
Rohan~Kumar Das, Tomi Kinnunen, Wen-Chin Huang, Zhenhua Ling, Junichi
  Yamagishi, Yi~Zhao, Xiaohai Tian, and Tomoki Toda,
\newblock ``{Predictions of subjective ratings and spoofing assessments of
  {Voice Conversion Challenge} 2020 submissions},''
\newblock in {\em Proc. Joint Workshop for the Blizzard Challenge and Voice
  Conversion Challenge 2020}, 2020, pp. 99--120.

\bibitem{watanabe2018espnet}
Shinji Watanabe, Takaaki Hori, Shigeki Karita, Tomoki Hayashi, Jiro Nishitoba,
  Yuya Unno, Nelson {Enrique Yalta Soplin}, Jahn Heymann, Matthew Wiesner,
  Nanxin Chen, Adithya Renduchintala, and Tsubasa Ochiai,
\newblock ``{ESPnet}: End-to-end speech processing toolkit,''
\newblock in {\em Proceedings of Interspeech}, 2018, pp. 2207--2211.

\bibitem{hayashi2019espnettts}
Tomoki Hayashi, Ryuichi Yamamoto, Katsuki Inoue, Takenori Yoshimura, Shinji
  Watanabe, Tomoki Toda, Kazuya Takeda, Yu~Zhang, and Xu~Tan,
\newblock ``{ESPnet-TTS}: Unified, reproducible, and integratable open source
  end-to-end text-to-speech toolkit,'' 2020.

\bibitem{WANG2020101114}
Xin Wang, Junichi Yamagishi, Massimiliano Todisco, H{\'{e}}ctor Delgado,
  Andreas Nautsch, Nicholas Evans, Md~Sahidullah, Ville Vestman, Tomi Kinnunen,
  Kong~Aik Lee, Lauri Juvela, Paavo Alku, Yu-Huai Peng, Hsin-Te Hwang, Yu~Tsao,
  Hsin-Min Wang, S{\'{e}}bastien~Le Maguer, Markus Becker, Fergus Henderson,
  Rob Clark, Yu~Zhang, Quan Wang, Ye~Jia, Kai Onuma, Koji Mushika, Takashi
  Kaneda, Yuan Jiang, Li-Juan Liu, Yi-Chiao Wu, Wen-Chin Huang, Tomoki Toda,
  Kou Tanaka, Hirokazu Kameoka, Ingmar Steiner, Driss Matrouf,
  Jean-Fran{\c{c}}ois Bonastre, Avashna Govender, Srikanth Ronanki, Jing-Xuan
  Zhang, and Zhen-Hua Ling,
\newblock ``{ASVspoof 2019: a large-scale public database of synthesized,
  converted and replayed speech},''
\newblock {\em Computer Speech {\&} Language}, p. 101114, 2020.

\bibitem{Todisco2019}
Massimiliano Todisco, Xin Wang, Ville Vestman, Md. Sahidullah, H{\'{e}}ctor
  Delgado, Andreas Nautsch, Junichi Yamagishi, Nicholas Evans, Tomi~H Kinnunen,
  and Kong~Aik Lee,
\newblock ``{ASVspoof 2019: future horizons in spoofed and fake audio
  detection},''
\newblock in {\em Proc. Interspeech}, 2019, pp. 1008--1012.

\bibitem{wu2019blizzard}
Zhizheng Wu, Zhihang Xie, and Simon King,
\newblock ``The blizzard challenge 2019,''
\newblock in {\em Proc. Blizzard Challenge Workshop}, 2019, vol. 2019.

\bibitem{wang2018comparison}
Xin Wang, Jaime Lorenzo-Trueba, Shinji Takaki, Lauri Juvela, and Junichi
  Yamagishi,
\newblock ``A comparison of recent waveform generation and acoustic modeling
  methods for neural-network-based speech synthesis,''
\newblock in {\em 2018 IEEE International Conference on Acoustics, Speech and
  Signal Processing (ICASSP)}. IEEE, 2018, pp. 4804--4808.

\bibitem{kawai2004ximera}
Hisashi Kawai, Tomoki Toda, Jinfu Ni, Minoru Tsuzaki, and Keiichi Tokuda,
\newblock ``{XIMERA: A new TTS from ATR} based on corpus-based technologies,''
\newblock in {\em Fifth ISCA Workshop on Speech Synthesis}, 2004.

\bibitem{yang2021superb}
Shu-wen Yang, Po-Han Chi, Yung-Sung Chuang, Cheng-I~Jeff Lai, Kushal Lakhotia,
  Yist~Y Lin, Andy~T Liu, Jiatong Shi, Xuankai Chang, Guan-Ting Lin, et~al.,
\newblock ``{SUPERB:} speech processing universal performance benchmark,''
\newblock {\em Interspeech}, 2021.

\bibitem{librispeech}
Vassil Panayotov, Guoguo Chen, Daniel Povey, and Sanjeev Khudanpur,
\newblock ``Librispeech: an asr corpus based on public domain audio books,''
\newblock in {\em 2015 IEEE international conference on acoustics, speech and
  signal processing (ICASSP)}. IEEE, 2015, pp. 5206--5210.

\bibitem{librilight}
Jacob Kahn, Morgane Rivi{\`e}re, Weiyi Zheng, Evgeny Kharitonov, Qiantong Xu,
  Pierre-Emmanuel Mazar{\'e}, Julien Karadayi, Vitaliy Liptchinsky, Ronan
  Collobert, Christian Fuegen, et~al.,
\newblock ``Libri-light: A benchmark for asr with limited or no supervision,''
\newblock in {\em ICASSP 2020-2020 IEEE International Conference on Acoustics,
  Speech and Signal Processing (ICASSP)}. IEEE, 2020, pp. 7669--7673.

\bibitem{commonvoice}
Rosana Ardila, Megan Branson, Kelly Davis, Michael Kohler, Josh Meyer, Michael
  Henretty, Reuben Morais, Lindsay Saunders, Francis Tyers, and Gregor Weber,
\newblock ``{Common Voice:} a massively-multilingual speech corpus,''
\newblock in {\em Proceedings of the 12th Language Resources and Evaluation
  Conference}, 2020, pp. 4218--4222.

\bibitem{switchboard}
John~J Godfrey, Edward~C Holliman, and Jane McDaniel,
\newblock ``{SWITCHBOARD:} telephone speech corpus for research and
  development,''
\newblock in {\em Acoustics, Speech, and Signal Processing, IEEE International
  Conference on}. IEEE Computer Society, 1992, vol.~1, pp. 517--520.

\bibitem{fisher}
Christopher Cieri, David Miller, and Kevin Walker,
\newblock ``{The Fisher Corpus:} a resource for the next generations of
  speech-to-text,''
\newblock in {\em Proceedings of the Fourth International Conference on
  Language Resources and Evaluation (LREC’04)}, 2004.

\bibitem{mls}
Vineel Pratap, Qiantong Xu, Anuroop Sriram, Gabriel Synnaeve, and Ronan
  Collobert,
\newblock ``{MLS:} a large-scale multilingual dataset for speech research,''
\newblock in {\em INTERSPEECH}, 2020.

\bibitem{babel}
Mary Harper,
\newblock ``{BABEL: IARPA} solicitation {IARPA-BAA-11-02},''
\newblock 2011.

\end{thebibliography}
}
\end{document}